# Comparing Complex Chemistry in Neighboring Hot Cores: NOEMA Studies of W3(H$_2$O) and W3(OH)

Will E. Thompson[1], Morgan M. Giese[2], Dariusz C. Lis[3], and Susanna L. Widicus Weaver[1,2]
[1] Department of Chemistry, University of Wisconsin–Madison, 1101 University Ave., Madison, WI 53706, USA
[2] Department of Astronomy, University of Wisconsin–Madison, 475 N. Charter St., Madison, WI 53706, USA
[3] Jet Propulsion Laboratory, California Institute of Technology, 4800 Oak Grove Drive, Pasadena, CA 91109, USA; slww@chem.wisc.edu
Received 2022 November 16; revised 2023 June 2; accepted 2023 June 2; published 2023 July 18

## Abstract

Presented here are NOEMA interferometric observations of the neighboring hot cores W3(H$_2$O) and W3(OH). The presence of two star-forming cores at different evolutionary stages within the same parent cloud presents a unique opportunity to study how the physics of the source and its evolutionary stage impact the chemistry. Through spectral analysis and imaging, we identify over 20 molecules in these cores. Most notably, we have detected HDO and CH$_3$CH$_2$CN in W3(OH), which were previously not detected in this core. We have imaged the molecular emission, revealing new structural features within these sources. W3(OH) shows absorption in a "dusty cocoon" surrounded by molecular emission. These observations also reveal extended emission that is potentially indicative of a low-velocity shock. From the information obtained herein, we have constructed column density and temperature maps for methanol and compared this information to the molecular images. By comparing the spatial distribution of molecules that may be destroyed at later stages of star formation, this work demonstrates the impact of physical environment on chemistry in star-forming regions at different evolutionary stages.

*Unified Astronomy Thesaurus concepts:* Astrochemistry (75); Interferometry (808); Star forming regions (1565); H II regions (694)

*Supporting material:* data behind figures

## 1. Introduction

The process of star formation, from the collapse of molecular clouds to the aggregation of icy planetesimals, establishes rich environments for complex chemistry to occur. The majority of this chemistry occurs on the icy mantles of dust grains, which form from the freezeout of molecules onto dust during the cold collapse phase of a molecular cloud (Blake & van Dishoeck 1998; van Dishoeck 2006; Herbst & van Dishoeck 2009). This ice chemistry is actively modified by cosmic rays, ultraviolet radiation, and thermal processing, which leads to the formation of larger organic molecules (Munoz Caro et al. 2002; Öberg et al. 2011; Altwegg et al. 2019).

The radiative processing of ice mantles has been shown to form prebiotic molecules, which are crucial for the origin of life. These prebiotic molecules have been routinely detected in extraterrestrial environments such as star-forming regions and comets (Blake & van Dishoeck 1998; Jørgensen et al. 2012; Widicus Weaver & Friedel 2012; Altwegg et al. 2019; Rivilla et al. 2020; Ligterink et al. 2022). This observed chemistry makes up the bulk chemical inventory from which a planetary system may form. It is therefore crucial to study molecular complexity at the early stages of star formation. The mechanisms for survival and delivery of these molecules during the process of planet formation still remain undefined (Blake & van Dishoeck 1998; van Dishoeck 2006). Nonetheless, such material could be delivered to a planet via meteorite or cometary impacts, seeding the formation of life (Oró 1961; Chyba et al. 1990; Chyba & Sagan 1992; Hartogh et al. 2011).

Rotational transitions of molecules are incredibly useful diagnostic tools for probing the physical characteristics of interstellar regions such as densities, temperatures, chemical abundances, and kinematics (Tychoniec et al. 2021). By using millimeter-wave interferometers with high spatial and spectral resolution to observe molecular lines, the physical environment of star-forming regions can be analyzed alongside the observed chemistry to better understand how chemical networks are influenced by different physical environments (Jørgensen et al. 2004; Tobin et al. 2011; Tychoniec et al. 2021).

Neighboring star-forming cores embedded in the same dense cloud are uniquely suited to study how molecular complexity may vary within different interstellar regions. In the work presented herein, millimeter-wave observations were conducted of the high-mass star-forming region W3, which consists of the neighboring cores W3(H$_2$O) and W3(OH), the latter being an ultracompact H II region. W3 is at a distance of ∼2 kpc from Earth with a $v_{lsr}$ of −47 km s$^{-1}$ (Rivera-Ingraham et al. 2013). The two cores are named as such owing to their association with H$_2$O and OH masers, respectively (Wynn-Williams et al. 1972; Wyrowski et al. 1997; Wilner et al. 1999).

W3(H$_2$O) has been shown to exhibit vastly different chemistry than W3(OH), likely due to being at a younger evolutionary stage (Helmich & van Dischoeck 1997; Qin et al. 2015; Widicus Weaver et al. 2017). This work presents imaging studies focusing on the complex organic chemistry of the W3 star-forming region, along with an analysis of the physical parameters associated with the detected molecules. Our goal is to study the chemical and physical differences between the neighboring cores W3(H$_2$O) and W3(OH). Such studies will enable us to form a multifaceted understanding of how prebiotic molecules are formed and distributed during star and planet formation.

**Table 1**
Summary of High-resolution Spectral Windows

| Spectral Window | Frequency Range (MHz) | rms (mJy beam$^{-1}$) | Beam Position Angle (deg) |
|---|---|---|---|
| 1 | 127,823–127,888 | 5.12 | −185.5 |
| 2 | 127,950–128,080 | 5.13 | −185.5 |
| 3 | 128,589–128,849 | 5.35 | −184.9 |
| 4 | 129,103–129,168 | 5.41 | −180.0 |
| 5 | 129,354–129,746 | 4.62 | −180.0 |
| 6 | 130,190–130,319 | 4.67 | −179.8 |
| 7 | 132,239–132,303 | 4.65 | −3.10 |
| 8 | 132,557–132,944 | 4.39 | −4.11 |
| 9 | 133,133–133,328 | 4.15 | −179.7 |
| 10 | 133,382–133,647 | 4.36 | −180.0 |
| 11 | 134,157–134,287 | 4.66 | −180.0 |
| 12 | 134,861–134,926 | 4.36 | −179.9 |
| 13 | 134,989–135,055 | 4.38 | −179.9 |
| 14 | 135,245–135,311 | 4.92 | −180.0 |
| 15 | 143,116–143,181 | 4.96 | −185.0 |
| 16 | 143,435–143,566 | 5.06 | −189.9 |
| 17 | 143,692–143,756 | 4.91 | −184.6 |
| 18 | 144,522–144,653 | 5.04 | −185.0 |
| 19 | 144,714–144,910 | 5.46 | −6.87 |
| 20 | 145,036–145,163 | 5.16 | −188.6 |
| 21 | 145,547–145,612 | 5.42 | −4.91 |
| 22 | 145,931–145,996 | 4.92 | 174.7 |
| 23 | 146,315–146,379 | 4.96 | −185.0 |
| 24 | 146,507–146,635 | 5.40 | −5.19 |
| 25 | 146,954–147,213 | 6.02 | −6.68 |
| 26 | 148,042–148,172 | 5.45 | −180.0 |
| 27 | 149,514–149,579 | 5.99 | −186.5 |
| 28 | 150,091–150,666 | 6.59 | −6.51 |

## 2. Observations and Data Reduction

W3 was observed with the IRAM/NOEMA interferometer[4] in the C and D configurations for ∼4 hr in each configuration on 2021 August 21 and 26 and 2021 October 19. The pointing center of the observations was $\alpha$(J2000) = 02$^h$27$^m$03$^s$.87, $\delta$ (J2000) = 61°52′24″.6, so that the field of view included emission from both the W3(H$_2$O) and W3(OH) cores. The observations were in the $\lambda$ = 2 mm wavelength range, centered at a local oscillator frequency of 145.255 GHz with frequency coverage from 127.823 to 135.311 GHz and from 143.116 to 150.666 GHz for the lower and upper sidebands, respectively.

The baselines of the array ranged from 24.0 to 176.0 m in the 10D configuration and from 24.0 to 368.0 m in the 10C configuration. The data were obtained using the PolyFiX correlator, which instantaneously processes ∼31 GHz of spectral bandwidth for each antenna in both sidebands. This spectral setup gave a channel spacing of 2.0 MHz across both sidebands. Additionally, 28 high-resolution spectral windows were selected within the two sidebands with channel spacings of 62.5 kHz. Bandpass calibration was performed using observations of 3C 84, and flux calibration was performed using observations of LKHA101 and 2010+723. Phase and amplitude calibration was performed using J0224+6228 and 0224+671. The system temperatures varied between 100 and 250 K during the observations. A summary of the observational setup is shown in Table 1.

The data sets were reduced using the GILDAS packages CLIC and MAPPING.[5] Self-calibration was performed on the data set by first creating a continuum image, constructed from line-free channels. This self-calibration solution was then applied to the 28 high-resolution spectral datacubes. The datacubes were cleaned using robust weighting until they converged to a value of twice the noise for the maximum amplitude of the absolute value of the residual. The background temperature was calculated based on the peak continuum flux for each core, giving 2.6 K for W3(H$_2$O) and 19.7 K for W3(OH). These background temperatures are less than 10% of the derived methanol temperature, which is a good proxy for the kinetic temperature since methanol has many transitions covering a wide range of energies. Furthermore, the spectral fitting shows that most lines are optically thin. We can therefore assume that the contribution from the background to the line intensities is not significant. A linear baseline was subtracted during the data reduction process so that the resultant spectra are centered at 0 Jy beam$^{-1}$. The resulting synthesized beams were ∼1″.87 × 1″.34 (PA = −9.3°) in the lower sideband and ∼1″.69 × 1″.22 (PA = −7.9°) in the upper sideband.

## 3. Continuum Imaging

The continuum was imaged at 132 GHz, as seen in Figure 1. At this frequency, the continuum emission from W3(OH) consists mostly of free–free emission, while that of W3(H$_2$O) is due to dust (Wilner et al. 1995; Wyrowski et al. 1997, 1999; Stecklum et al. 2002). The rms of the continuum was determined from an emission-free region in the image. A two-dimensional Gaussian was fit to the continuum for both W3(H$_2$O) and W3(OH) using GILDAS. The continuum emission has a fitted source size of 2″.47 × 2″.66 at 134° (from north toward east) for W3(H$_2$O) and 2″.20 × 2″.42 at 85°.1 for W3(OH). These source sizes are larger than the synthesized beam size for the observations, indicating that the emission is spatially resolved.

## 4. Line Identification and Analysis

In order to analyze the broadband spectra, the Global Optimization and Broadband Analysis Software for Interstellar Chemistry (GOBASIC) was used. The intricacies of this software are described by Rad et al. (2016). GOBASIC assumes both local thermodynamic equilibrium (LTE) and Gaussian line shapes. Using these assumptions, GOBASIC performs a broadband spectral analysis for multiple molecules and components in order to derive column density in square centimeters, spectral line FWHM ($\Delta v$) in kilometers per second, temperature in kelvin, and velocity shift relative to the $v_{lsr}$ of $-47$ km s$^{-1}$ in kilometers per second for each component.

The analysis and fitting procedure used herein is similar to that presented in a previous published work using GOBASIC (Widicus Weaver et al. 2017). In short, the molecular catalog information is loaded into GOBASIC along with initial guesses for temperature, column density, line width, and shift relative to the LSR velocity. The spectral simulation is then compared to the observational data set using the pattern search algorithm until the best match is achieved and the physical parameters are optimized. The spectrum is simulated using the catalog line frequencies and intensity information, as well as the partition function information based on the catalog datasheet. The partition function is interpolated to a temperature equal to the determined rotational temperature using the general functional form

$$Q(T) = \alpha T^\beta [\gamma + \exp(\epsilon/T)]. \quad (1)$$

---
[4] IRAM is supported by INSU/CNRS (France), MPG (Germany), and IGN (Spain).

[5] http://www.iram.fr/IRAMFR/GILDAS

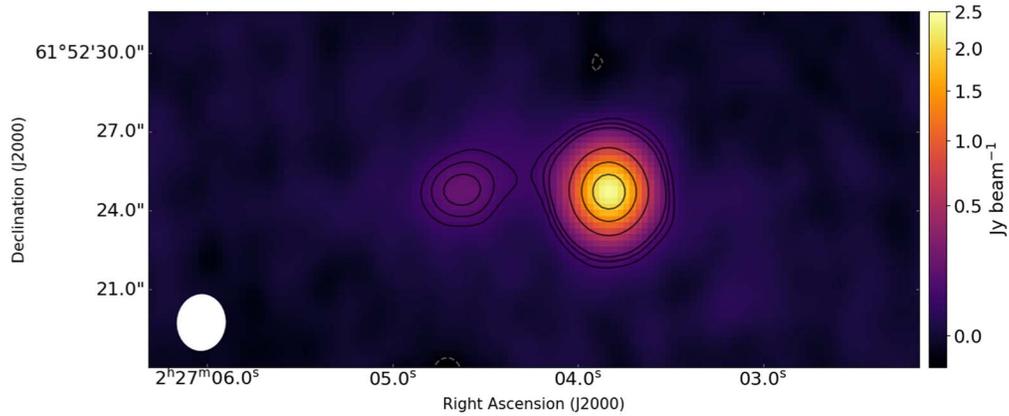

**Figure 1.** 132 GHz continuum emission of W3(H$_2$O) (left) and W3(OH) (right). Black contour levels are at 12, 18, 25, 100, 200, 300 × $\sigma$ ($\sigma$ = 6.481 mJy beam$^{-1}$). The gray dotted contour level is at $-3\sigma$. The synthesized beam size is shown in the lower left corner. The color bar is normalized using a square root stretch to show the large differences in emission between W3(H$_2$O) and W3(OH).

The coefficients for the partition function interpolation for each molecule are summarized in a previous work (Widicus Weaver et al. 2017). The parameters for each fit were constrained within a reasonable limit for the search grid, where the column density, line width, temperature, and velocity shift were held to a range of physically meaningful values. Given that one of the major goals of the current high spatial resolution imaging work is to determine the source morphology for molecular emission for each detected molecule, no corrections for source size were included in this analysis. Therefore, the emission is assumed to be resolved and to fill the small synthesized beam (see discussion below).

We used published line surveys of W3(H$_2$O) and W3(OH) to determine the initial set of molecules to analyze in this data set (Helmich et al. 1994; Helmich & van Dischoeck 1997; Fontani et al. 2007; Hernández-Hernández et al. 2014; Qin et al. 2015, 2016; Widicus Weaver et al. 2017; Ahmadi et al. 2018). Additional molecules were added to the analysis based on spectral lines that were observed and their match to other species found in the catalogs. All spectral line catalogs were imported from the Cologne Database for Molecular Spectroscopy and the Jet Propulsion Laboratory (JPL) Spectral Line Catalog (Müller et al. 2001, 2005; Pickett et al. 1998).

Spectra were extracted for W3(H$_2$O) and W3(OH) from individual pixels in the datacubes. The lower-sideband and upper-sideband spectra for each core can be seen in Figure 2. The resulting spectra had an average rms of 0.0629 K in a 4.31 km s$^{-1}$ velocity channel. For both W3(H$_2$O) and W3(OH), these spectra were from the center of the core at the pixel corresponding to the peak of the continuum flux. For W3(H$_2$O), this corresponds to a R.A. of 2$^h$27$^m$04$^s$.6 and a decl. of 61°52′25″.0 (J2000). For W3(OH), this corresponds to a R.A. of 2$^h$27$^m$03$^s$.8 and a decl. of 61°52′25″.3 (J2000).

Numerous molecular species display transitions in the frequency range covered by our observations. Detections were obtained for CH$_3$CH$_2$CN (ethyl cyanide), CH$_3$CN (methyl cyanide), CH$_3$OCH$_3$ (dimethyl ether), CH$_3$OH (methanol), HCOOCH$_3$ (methyl formate), and SO$_2$ (sulfur dioxide) in both W3(H$_2$O) and W3(OH). This includes the first detection of CH$_3$CH$_2$CN in W3(OH), as previous studies did not detect emission from this molecule (Qin et al. 2015). The GOBASIC results for W3(OH) presented in Figure 3 reveal that the spectral features for CH$_3$CH$_2$CN are severely blended with lines arising from other molecules in this source. Unfortunately, no obvious identification of the carriers for these lines can be made based on the information in publicly available spectral line catalogs. Nonetheless, there are numerous spectral features for CH$_3$CH$_2$CN that are clearly observed with high signal-to-noise ratios in these spectral windows, despite the blending. Due to this blending, it was difficult to derive physically meaningful parameters for CH$_3$CH$_2$CN in W3(OH).

SiO (silicon monoxide), SO (sulfur monoxide), $^{33}$SO, $^{34}$SO, HCS$^+$ (thioformyl cation), HNCO (isocyanic acid), H$_2$CS (thioformaldehyde), OCS (carbonyl sulfide), H$_2$CO (formaldehyde), NH$_2$CHO (formamide), HDO (deuterated water), CS (carbon monosulfide), C$^{33}$S, C$^{34}$S, H$_2^{13}$CO, NO (nitric oxide), DCN (deuterium cyanide), HC$_3$N (cyanoacetylene), HC$_3$N $v_7$, $^{33}$SO$_2$, and $^{34}$SO$_2$ were also detected. However, for these molecules only a few lines are observable in these frequency windows and their physical parameters could not be accurately determined.

Table 2 shows the derived physical parameters for the molecules detected in the two cores. Included here are the column density, temperature, line width, and velocity shift relative to the $v_{lsr}$ used for the observations. For each molecule fit in Table 2, vibrational corrections to the partition function were not applied. The $v = 0$ states were fit for SO$_2$, CH$_3$CN, CH$_3$CH$_2$CN, and CH$_3$OCH$_3$; $v_t = 0$–2 for CH$_3$OH; and $v_t = 0$–1 for HCOOCH$_3$. To check that no beam filling factor was required, an additional fit for the same molecules in W3(OH) was performed using a source size of 2″.0 as determined for CH$_3$OCH$_3$, which displays the most compact emission. Details on the calculation of this source size can be found in Section 5. In this secondary fit, the column density of CH$_3$OCH$_3$ is 8.1e+16 cm$^{-2}$ with an uncertainty of 1.4e+16 cm$^{-2}$. This value is 1.4 times higher than the value computed without correction for beam dilution and reported in Table 2. This is less than a factor of two, implying that any difference factor for other molecules with more extended

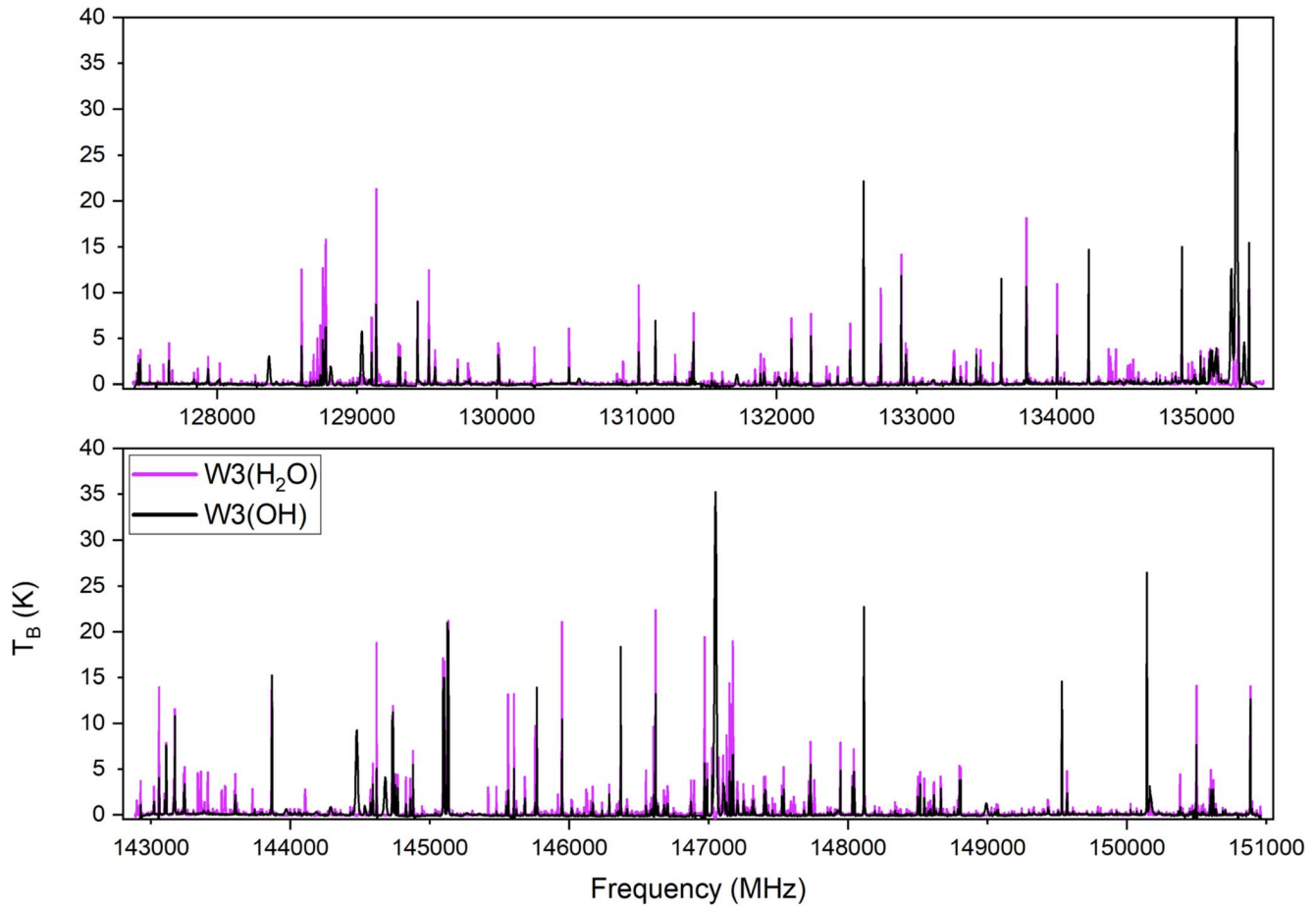

**Figure 2.** Spectra of W3(H$_2$O) and W3(OH) for the lower sideband (top) and upper sideband (bottom). The data behind the figure are available as supplementary information.

(The data used to create this figure are available.)

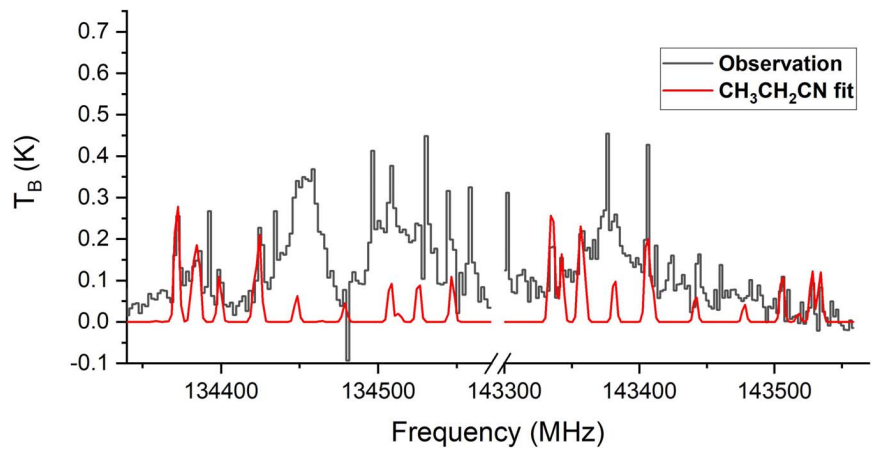

**Figure 3.** GOBASIC fit (red) of CH$_3$CH$_2$CN in the observational data (black) for both sidebands for W3(OH). The GOBASIC data behind the figure for six molecules for W3 (H$_2$O) and five molecules for W3(OH) are available as supplementary information. These data can be used to derive the parameters in Table 2.

(The data used to create this figure are available.)

emission would be even smaller. Therefore, the assumption to neglect beam dilution in these calculations is reasonable.

Comparing the physical parameters calculated by GOBASIC between the two cores reveals that all of the identified species are observed with similar line widths and velocity shifts. However, the line widths for CH$_3$CN and SO$_2$ in W3(OH) observations are not well determined, having 1$\sigma$ uncertainties of 1.00 and 2.44 km s$^{-1}$, respectively. While the cores have different shifts relative to the LSR velocity, the shifts are similar for all molecules detected in a given core. The line

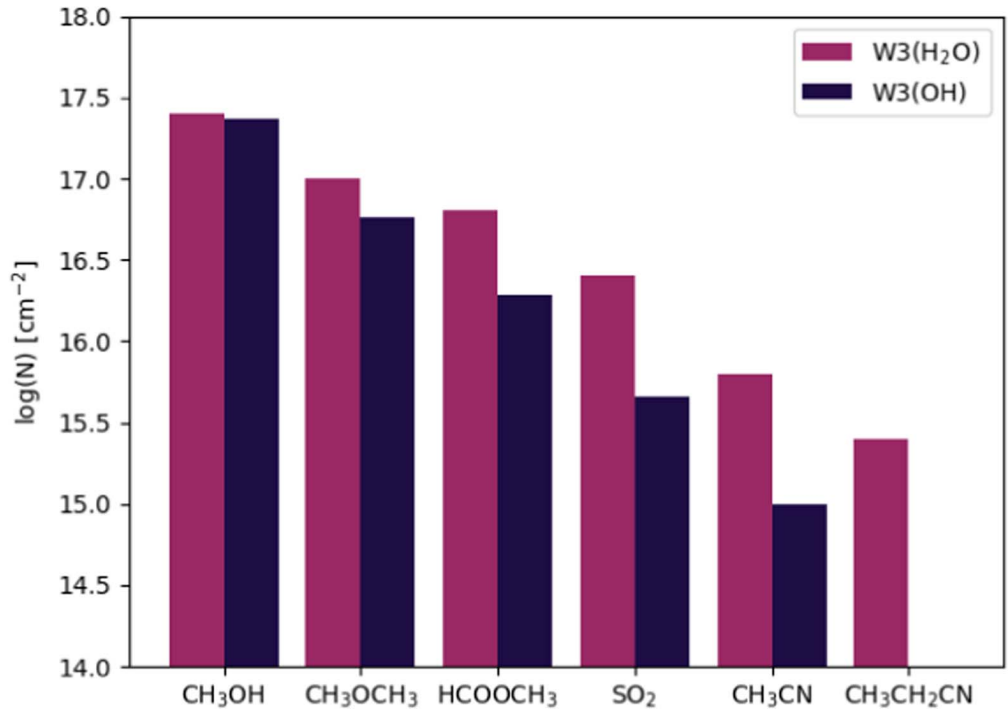

**Figure 4.** Comparison of the derived column densities for molecules detected in W3($H_2O$) and W3(OH).

Table 2
Calculated Physical Parameters for W3($H_2O$) and W3(OH) from GOBASIC Fits

| Molecule | $\log_{10} N_T$ | $N_T$ (cm$^{-2}$) | Temperature (K) | FWHM (km s$^{-1}$) | Shift (km s$^{-1}$) |
|---|---|---|---|---|---|
| | | W3($H_2O$) Derived Parameters | | | |
| $CH_3OH$ | 17.38(1) | 2.411(31)e+17 | 187(2) | 7.18(5) | −2.96(3) |
| $CH_3OCH_3$ | 17.03(1) | 1.066(29)e+17 | 108(3) | 6.72(16) | −2.76(9) |
| $HCOOCH_3$ | 16.83(2) | 6.700(325)e+16 | 164(6) | 6.81(11) | −2.91(6) |
| $CH_3CN$ | 15.76(3) | 5.751(36)e+15 | 191(10) | 8.66(13) | −2.38(6) |
| $CH_3CH_2CN$ | 15.41(4) | 2.590(249)e+15 | 153(16) | 8.27(26) | −1.46(13) |
| $SO_2$ | 16.35(1) | 2.237(53)e+16 | 104(3) | 9.72(13) | −2.27(7) |
| | | W3(OH) Derived Parameters | | | |
| $CH_3OH$ | 17.37(3) | 2.360(146)e+17 | 209(8) | 5.51(19) | 1.83(8) |
| $CH_3OCH_3$ | 16.76(8) | 5.722(1021)e+16 | 122(19) | 6.09(91) | 1.45(62) |
| $HCOOCH_3$ | 16.29(12) | 1.938(554)e+16 | 108(25) | 4.83(72) | 1.75(29) |
| $CH_3CN$ | 15.00(15) | 1.005(339)e+15 | 126(39) | 6.06(100) | 1.98(47) |
| $SO_2$ | 15.66(15) | 4.547(1533)e+15 | 80(29) | 5.91(244) | 1.23(46) |

**Note.** The five parameters are the base-10 log of column density ($\log_{10} N_T$), column density in square centimeters ($N_T$), temperature in kelvin, spectral line FWHM in kilometers per second, and velocity shift relative to the $v_{lsr}$ in kilometers per second (Shift). The uncertainties (written in parentheses) are expressed in units of the last digit.

widths between the two cores are generally comparable but are larger in W3($H_2O$). Molecular column densities in the two cores are also comparable to the values reported in the literature for these sources (Qin et al. 2015). Figure 4 shows a comparison between the column densities of these molecules in each core.

We find $CH_3OCH_3$ to have a lower column density when compared to $CH_3OH$, due to $CH_3OCH_3$ likely being formed from $CH_3OH$ photolysis on ice grain mantles as described in the literature (Garrod & Herbst 2006; Garrod et al. 2008, 2022). In both sources, we observe that $CH_3OCH_3$ has a higher column density than $HCOOCH_3$. Comparing the temperatures between the two cores, $HCOOCH_3$ and $CH_3CN$ are warmer in W3($H_2O$), while $CH_3OH$ is warmer in W3(OH). $CH_3OCH_3$ and $SO_2$ are generally comparable within the uncertainties in the two cores. In order to correctly derive the temperature values for $CH_3CN$ in W3(OH), the $8_1$–$7_1$ transition was flagged and excluded from the fit because it was severely blended with a hydrogen recombination line. In W3($H_2O$), a method similar to the one in Ahmadi et al. (2018) was used for $CH_3CN$ in which low-K transitions were flagged because they are optically thick. Using this method, we have observed the temperature of $CH_3CN$ to be comparable to that of $CH_3OH$ in W3($H_2O$) with both temperatures around 190 K. These results for the temperature of $CH_3CN$ agree with those calculated by Ahmadi et al. (2018).

### 4.1. Column Density and Temperature Maps

Given the high spatial resolution, the NOEMA spectral cubes can be used to create maps of temperature and column density for some molecules. CH$_3$OH serves well as a tool for such imaging, as it has a plethora of spectral lines that are bright and for which the relative intensities are highly dependent on temperature. To perform this analysis, the existing GOBASIC code was modified to accept a datacube as an input, extract spectra within a specified pixel range, baseline-correct each spectrum, and perform a spectral fit for a given molecule.

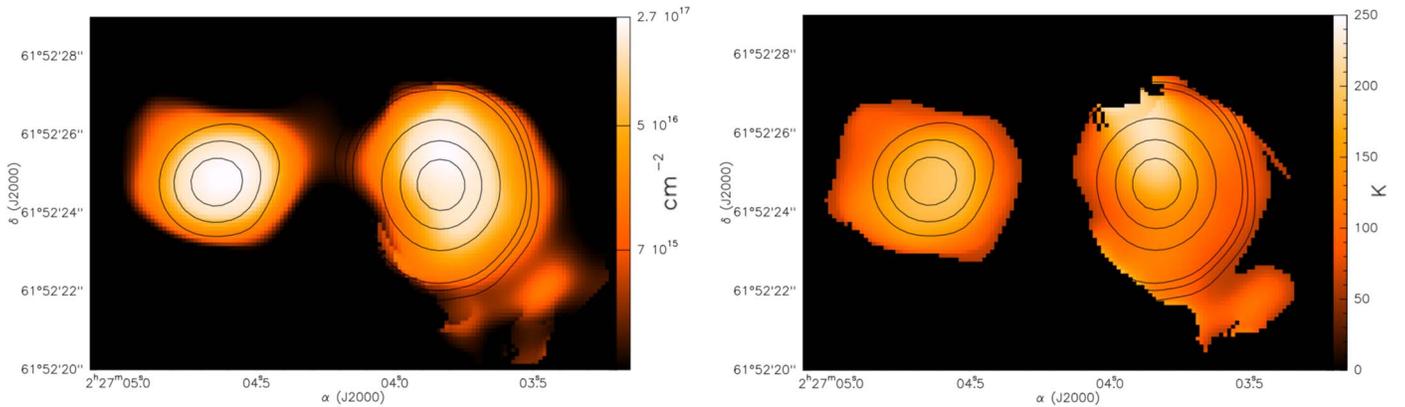

**Figure 5.** Column density (left) and temperature (right) maps for methanol. Black contour levels are the continuum at 12, 18, 25, 100, 200, 300 × $\sigma$ ($\sigma$ = 6.481 mJy beam$^{-1}$).

A fit was conducted using GOBASIC to determine the column density and temperature for CH$_3$OH using the spectra extracted at each pixel from the datacubes for the lower and upper sidebands. For each pixel, the corresponding spectra from each sideband were combined to create a single data file, with the channels between the two sidebands set to zero. Both a column density map and a temperature map were then constructed from this information, with the zeroed channels flagged so that they were not included in the fit. Furthermore, any pixels that resulted in poorly fit spectra (i.e., with large relative uncertainties) were flagged and set to a value of zero. The maps were then overlaid with the continuum image contours using GILDAS. The column density map was normalized using an equalization scaling in GILDAS. This scaling increases the visibility of low-contrast images by attempting to utilize the entirety of the chosen color map. The column density and temperature maps are shown in Figure 5.

This analysis shows that CH$_3$OH has similar column densities and temperatures in both cores. While the temperatures are similar between the cores, W3(OH) seems to have slightly higher temperatures than W3(H$_2$O) with temperatures slightly above and below 200 K, respectively. For W3(H$_2$O), the peak in column density and temperature for CH$_3$OH roughly coincides with the peak of the continuum emission. For W3(OH), the peak in column density and temperature for CH$_3$OH are offset from the continuum peak by $\sim$1″.

## 5. Imaging of Spectral Lines

The observations allowed emission from both W3(H$_2$O) and W3(OH) to be imaged using the 28 high-resolution spectral windows. Spectral lines covering a range of transitions were chosen to be imaged from the lower and upper sidebands. Using a Python script and the Astropy[6] Python library, moment-0 maps of the molecular emission were made by integrating over channels above 3$\sigma$, where $\sigma$ for each spectral window is given in Table 1 (Astropy Collaboration et al. 2013, 2018, 2022). All spectral lines that were imaged were carefully checked for blending with other spectral lines to avoid contamination in the moment-0 maps. Several transitions such as the H$_2$CS 4$_{1,4}$–3$_{1,3}$, CH$_3$CN 8$_1$–7$_1$, and NO 3/2–1/2 were blended with recombination lines and are thus excluded from our analysis. For HDO and DCN, only one transition was clearly detected. Both of these molecules have been detected in other observational studies of W3 (Helmich & van Dischoeck 1997; Qin et al. 2015). Additionally, the observed transitions are free of blending and are not contaminated with other molecules. Based on these reasons, we conclude that the detections are robust.

While most of the molecules detected in the line analysis were imaged, this was not possible for molecules with weak emission. Additionally, some of the detected molecules, such as NH$_2$CHO, were outside of the high-resolution spectral windows and therefore not imaged. The imaging results are presented in Figures 6 and 7 and discussed below.

A two-dimensional Gaussian was fit to the CH$_3$OCH$_3$ 6$_{3,3}$–6$_{2,4}$ transition, which has the most compact emission structure of any well-fitted molecule. CH$_3$OCH$_3$ was calculated to have a deconvolved source size of 2″.076 × 1″.889 at 64°.4 with an uncertainty of 0″.013 × 0″.013 for W3(H$_2$O) and 2″.825 × 2″.357 at 118°.3 with an uncertainty of 0″.031 × 0″.004 for W3(OH). This is more compact than the continuum but still spatially resolved by the synthesized beam of these observations. Since CH$_3$OCH$_3$ displays some of the most compact emission that we observed, all other molecules are also resolved by the synthesized beam. This calculated Gaussian fit value for CH$_3$OCH$_3$ in W3(OH) was used alongside the synthesized beam size of 1″.87 × 1″.34 to calculate the deconvolved source size of 2″.0 described in Section 4. To further verify these results, we performed a similar analysis for CH$_3$OH 5$_{-2}$–6$_{-1}$, which resulted in a calculated deconvolved source size of 2″.53 × 2″.36 at 81°.2 with an uncertainty of 0″.004 × 0″.003 for W3(H$_2$O) and 3″.50 × 2″.96 at 135°.0 with an uncertainty of 0″.028 × 0″.023 for W3(OH). All position angles listed here are defined from north toward east.

### 5.1. Cross-correlation of Moment-0 Maps

In order to quantitatively compare spatial distributions of molecules, cross-correlation of the moment-0 maps was calculated following a method similar to that used by Guzmán et al. (2018) and Law et al. (2021). The Pearson cross-correlation method quantitatively calculates the similarity of moment-0 maps by measuring the linear correlation between the two images. The more similar the spatial distributions are for emission in a pair of moment-0 maps, the closer their correlation coefficient is to 1.0.

---

[6] http://www.astropy.org

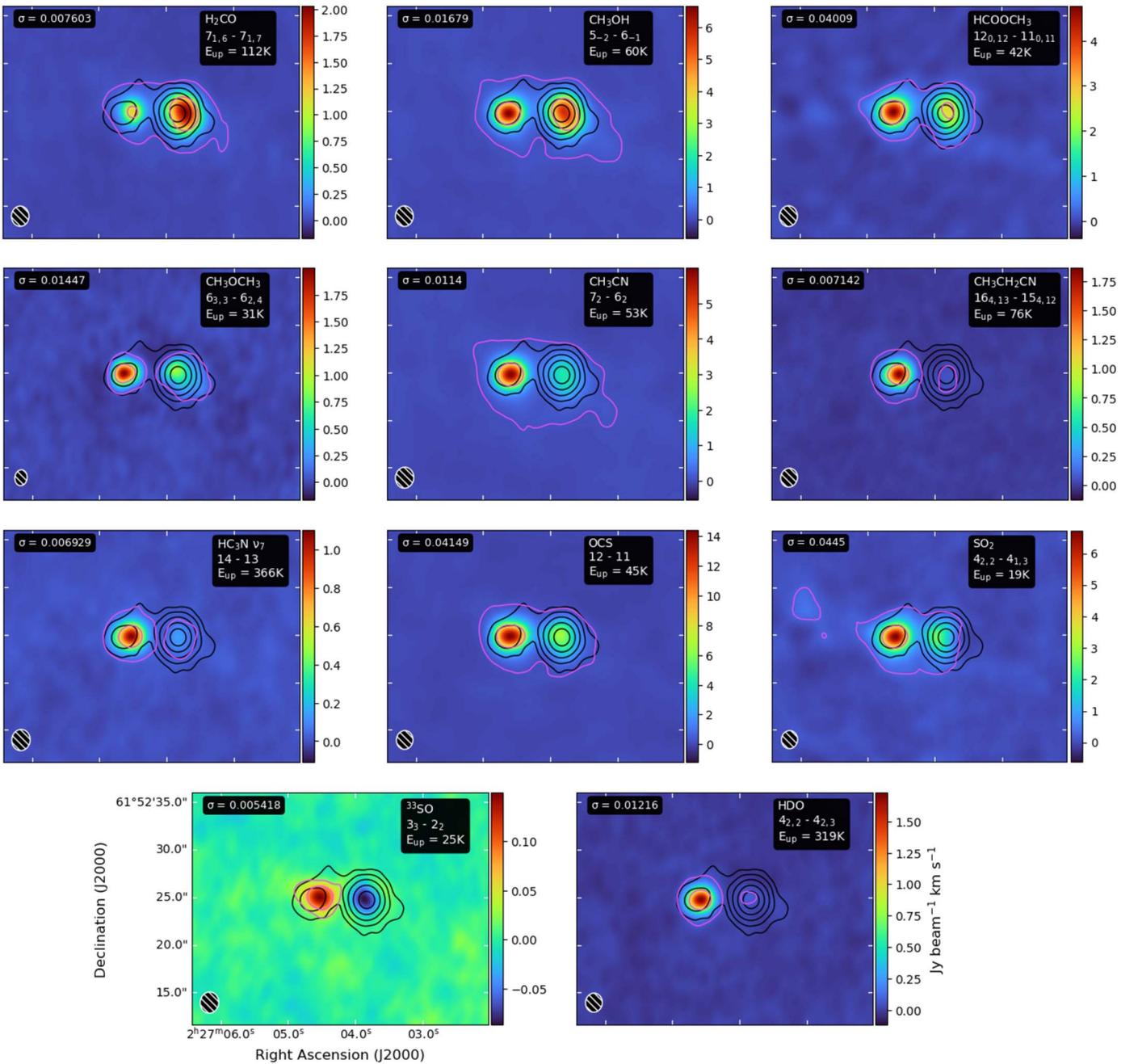

**Figure 6.** Imaged transitions showing compact emission. The integrated intensity is given by the color bar in each image in Jy beam$^{-1}$ km s$^{-1}$. Pink contours show $5 \times \sigma$ and 50% of the peak emission. Black contour levels for the continuum are at 5, 15, 45, 120, and $250 \times \sigma$ ($\sigma = 6.481$ mJy beam$^{-1}$). The noise levels for each map are shown in the upper left corner in units of Jy beam$^{-1}$ km s$^{-1}$. Beam size is shown in the lower left corner of each image.

Because all moment-0 maps contained emission from W3($H_2O$), a mask was applied to avoid the correlation being dominated by this source, similar to that done by Guzmán et al. (2018). The mask covering W3($H_2O$) is a circle 19 pixels in diameter, corresponding to a $4\rlap{.}{''}40$ diameter circle covering W3($H_2O$). Figure 8 displays the correlation coefficients for each pair of moment-0 maps.

While a high correlation coefficient between two molecules shows a degree of similarity in the morphology of their emission, it does not necessarily imply a chemical link between molecules. For molecules with both compact and extended emission morphology ($H_2CO$, $CH_3OH$), not every moment-0 map was cross-correlated. For $H_2CO$, the extended emission moment-0 map was used with the $2_{0,2}-1_{0,1}$ transition. For $CH_3OH$, the compact emission moment-0 map for the $5_{-2}-6_{-1}$ transition was used. The correlations of molecules will be discussed in Section 6.

## 6. Discussion

### 6.1. Compact Emission

Several molecules were observed to have compact emission that is cospatial with the continuum in these cores. Molecules displaying compact emission include $H_2CO$, $CH_3OH$, $CH_3OCH_3$, $HCOOCH_3$, $CH_3CN$, $CH_3CH_2CN$, $HC_3N$ $v_7$, $^{33}SO$, OCS, $SO_2$, and HDO.

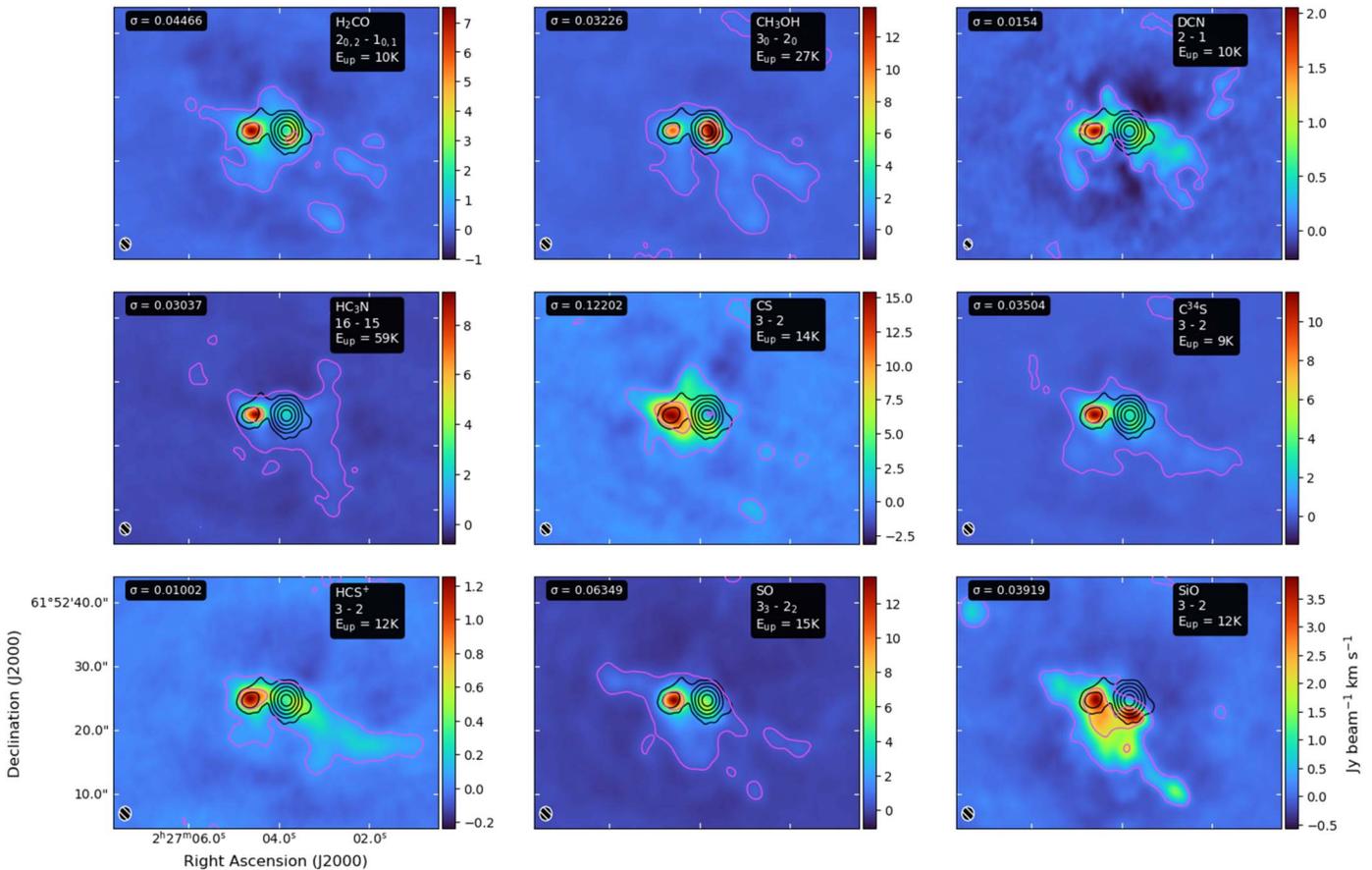

**Figure 7.** Imaged transitions showing extended emission. The integrated intensity is given by the color bar in each image in Jy beam$^{-1}$ km s$^{-1}$. Pink contours show $5 \times \sigma$ and 50% of the peak emission. Black contour levels for the continuum are at 5, 15, 45, 120, and $250 \times \sigma$ ($\sigma = 6.481$ mJy beam$^{-1}$). The noise levels for each map are shown in the upper left corner in units of Jy beam$^{-1}$ km s$^{-1}$. Beam size is shown in the lower left corner of each image.

This compact emission indicates that these molecules are either forming via gas-phase reactions in the hot core region or are being liberated into the gas phase when high temperatures are achieved. $H_2CO$ and $CH_3OH$ moment-0 maps of lines that cover different excitation energies show that the compact structure is excitation dependent. At an excitation energy above 60 K, the emission of $CH_3OH$ shows only a compact structure.

For $HCOOCH_3$ and $CH_3OCH_3$, the compact structure does not appear to be dependent on excitation energy. These two species are most often detected together and tend to trace the same spatial scales. Both molecules are likely formed from grain surface ice chemistry involving the $CH_3O$ radical, which is produced via methanol photodissociation (Garrod & Herbst 2006; Garrod et al. 2008; Laas et al. 2011; Garrod et al. 2022). Our observations of $CH_3OCH_3$ and $HCOOCH_3$ agree with these previously observed spatial trends, with $CH_3OCH_3$, $HCOOCH_3$, and $CH_3OH$ all showing high correlations of 0.91–0.97. Due to the formation processes of $CH_3OCH_3$ and $HCOOCH_3$, we would expect these to show similar spatial emission.

The sulfur-bearing molecules with compact emission are OCS, $^{33}SO$, and $SO_2$. $^{33}SO$ is seen almost exclusively in absorption in W3(OH). The presence of OCS in the hot core is expected, as it commonly traces ice sublimation and high-temperature chemistry (Tychoniec et al. 2021). The compact structure of $SO_2$ appears to not be dependent on excitation energy, with both the low- and higher-excitation energy transitions seen to have compact emission. The $SO_2$ $4_{2,2}$–$4_{1,3}$ transition shows an interesting structure with emission mostly seen toward the eastern side of W3(OH). Similar emission structure was seen by Wyrowski et al. 1999. OCS is the most highly correlated with the compact organics $CH_3OH$, $CH_3CN$, $CH_3OCH_3$, and $HCOOCH_3$. This trend was also seen by Guzmán et al. (2018). $^{33}SO$ was not cross-correlated with other molecules because the emission only appeared near the masked region.

Nitrogen-bearing molecules $CH_3CN$ and $CH_3CH_2CN$ are both well-known tracers of hot cores (Friedel & Widicus Weaver 2012; Widicus Weaver & Friedel 2012). Qin et al. (2015) find that $CH_3CN$ is most likely synthesized by high-temperature gas-phase reactions in $W3(H_2O)$, agreeing with the Orion observations by Widicus Weaver & Friedel (2012). However, Garrod et al. (2022) show that $CH_3CN$ is predominately formed in ice mantles during the early stages of star formation and then dominantly formed through gas-phase reactions during later stages. For $CH_3CH_2CN$, there is a lack of efficient formation pathways at later stages of star formation, causing it to undergo net destruction (Garrod et al. 2022). The moment-0 maps of $CH_3CH_2CN$ agree with these findings, where the low abundance of $CH_3CH_2CN$ in W3(OH) likely indicates a destruction pathway for this molecule at later stages of star formation, as W3(OH) is a more evolved H II region than the $W3(H_2O)$ core. The highest correlations for $CH_3CN$ are other compact organic molecules such as $CH_3OH$ and $HCOOCH_3$ and the sulfur-bearing molecule OCS. $CH_3CH_2CN$ is best correlated with HDO, although it is still a relatively low correlation of 0.62. Both of these molecules display similar moment-0 maps with low-intensity emission from W3(OH), but this does not indicate a chemical link between the two molecules.

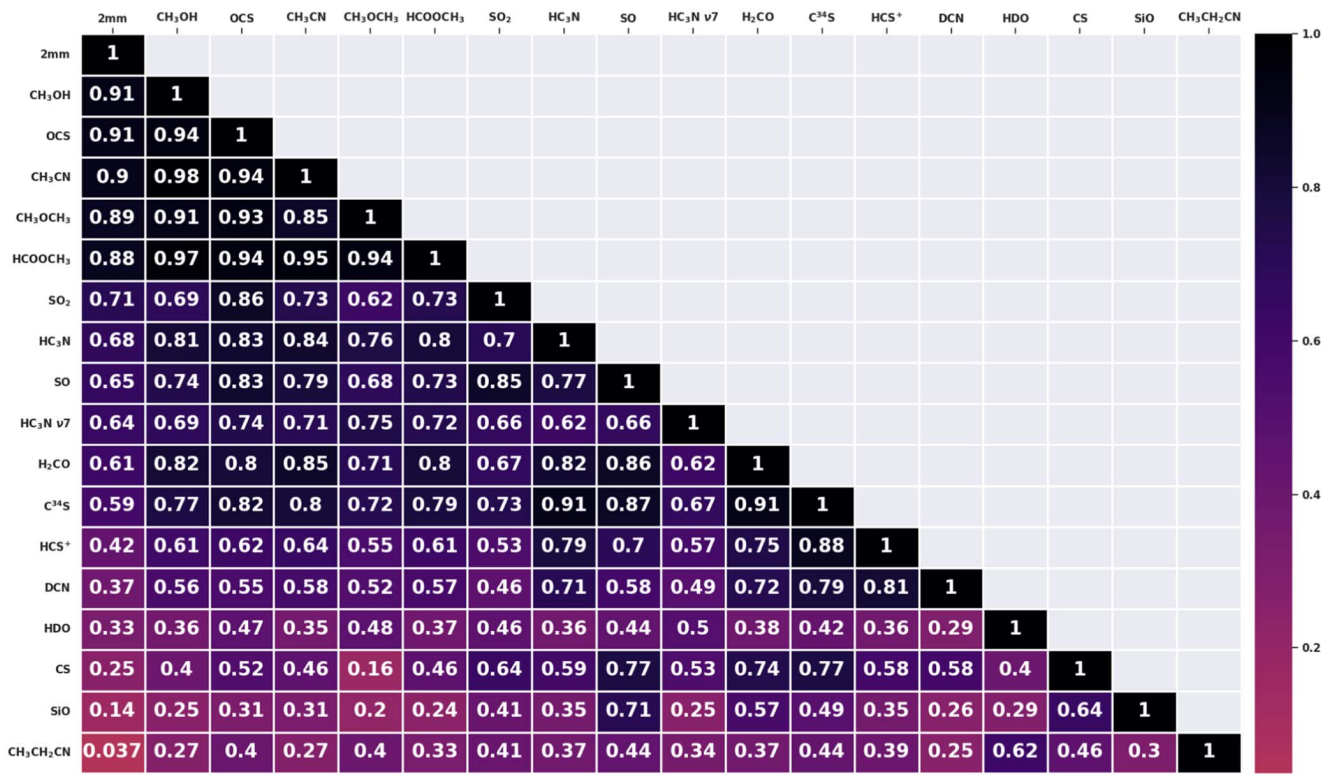

**Figure 8.** Correlation matrix of the imaged moment-0 maps shown in Figures 6 and 7. Correlation coefficient of 1.0 denotes maximum correlation. The "2mm" label denotes the 132 GHz continuum.

$HC_3N$ $v_7$ has compact emission and can be seen in both cores, though it is more abundant in $W3(H_2O)$. $HC_3N$ $v_7$ is most highly correlated to $CH_3OCH_3$ and OCS. This is not likely due to a correlation in chemical pathway, but rather because these molecules are commonly found to trace the warm, inner regions of hot cores (Rodríguez-Franco et al. 1998; Garrod & Herbst 2006; Tychoniec et al. 2021).

HDO was detected in both $W3(H_2O)$ and W3(OH), contrary to the findings of Qin et al. (2015). Based on their calculated ratio of $HDO/H_2O$, they predict that $W3(H_2O)$ is at an early evolutionary stage of star formation (Qin et al. 2015). The compact structure of HDO is quite different from the extended emission of the other deuterated molecule we detected, DCN. These selected transitions of HDO and DCN have a widely different upper-level energy (319 K for HDO versus 10 K for DCN). Therefore, the data at hand do not allow for a robust comparison of the spatial distribution of these molecules. HDO is likely tracing ice sublimation within the hot cores, while DCN can form through "warm" gas-phase chemistry from $CH_2D^+$ (Parise et al. 2009). Although deuterated molecules commonly trace cold gas, the observed transition of HDO has an upper-state energy of ∼319 K and is therefore only likely to be detected in the hot core. HDO is most highly correlated to $CH_3CH_2CN$. Similar to $CH_3CH_2CN$, it appears that the low abundance of HDO in W3(OH) agrees with destruction of this molecule in H II regions (Elitzur & de Jong 1978).

### 6.2. Extended Emission

Many of the moment-0 maps also show extended emission. The molecules with extended emission include DCN, SiO, CS, $C^{34}S$, $HCS^+$, SO, $H_2CO$, $CH_3OH$, and $HC_3N$.

SiO traces the bipolar molecular outflows observed by Zapata et al. (2011) in $W3(H_2O)$. The SiO moment-0 map traces both blueshifted and redshifted outflows toward the southwest and northeast, respectively, which agrees with the previous observations by Zapata et al. (2011). SiO shows maximum correlation to SO and CS. A high correlation between SiO and SO was also observed by Guzmán et al. (2018), indicating the association of SO with shocks.

In addition to emission correlating with the outflows traced by SiO, additional extended emission was observed that has previously only been imaged in $HCO^+$ and $CH_3OH$ emission (Sutton et al. 2004; Hakobian & Crutcher 2011). To the best of our knowledge, this is the first time this extended emission structure has been imaged for the molecules CS, $C^{34}S$, $HCS^+$, SO, DCN, and $H_2CO$. This structure is seen west of W3(OH) and does not trace the bipolar outflows of $W3(H_2O)$ as SiO does. The emission from each of these molecules appears to follow the same extended structure. The additional extended emission could be explained by the low energies of the transitions observed in this region. In the cases of CS, $C^{34}S$, $HCS^+$, SO, DCN, $H_2CO$, and $CH_3OH$, the imaged transitions have upper-state energies ranging from 9 to 27 K.

Alternatively, it is also possible that this emission is arising from molecules liberated by a low-velocity shock (i.e., 5 km s$^{-1} \leqslant \nu_s \leqslant 40$ km s$^{-1}$). Pineau des Forêts et al. (1992) and

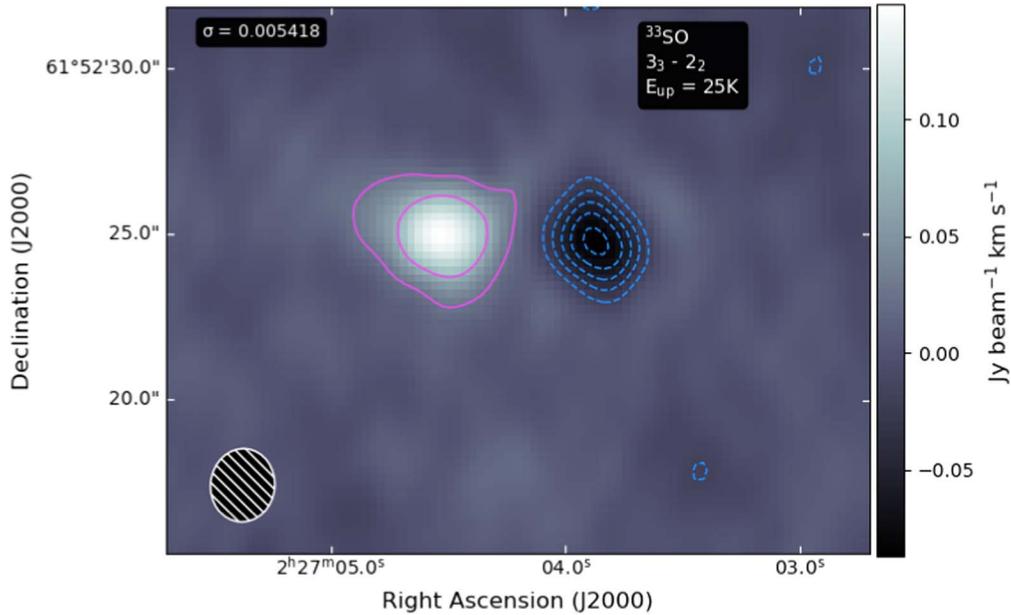

**Figure 9.** Imaged $^{33}$SO $3_3$–$2_2$ transition showing absorption in W3(OH). Pink contours show $5 \times \sigma$ and 50% of the peak emission. Blue contours show $-15, -12, -9, -6$, and $-3 \times \sigma$. The noise level is shown in the upper left corner in units of Jy beam$^{-1}$ km s$^{-1}$. Beam size is shown in the lower left corner.

Guzmán et al. (2018) explain how SO may be enhanced in low-velocity shocks when compared to SO$_2$. This is compatible with the moment-0 maps of SO$_2$, which only show compact emission structure at both low and high upper-state energies, while the SO moment-0 map shows extended emission. According to Guzmán et al. (2018), these low-velocity shocks can increase the amount of S$^+$ owing to production of secondary UV photons associated with the shock, accelerating the rate of ion−neutral reactions such as

$$S^+ + CH_2 \longrightarrow HCS^+ + H. \quad (2)$$

This is consistent with the moment-0 map of HCS+, which has extended emission only appearing west of W3(OH), where the proposed low-velocity shocked gas is located.

Even though CS is most highly correlated to SO and C$^{34}$S, the correlation to C$^{34}$S is still lower than one might expect with a coefficient of 0.77. Similar trends were seen by Guzmán et al. (2018) with CS and C$^{33}$S, which can be attributed to the high optical depth of CS. The majority of CS emission is seen near the continuum; however, there is $5\sigma$ emission southwest of W3(OH) and near the bipolar outflow traced by SiO.

For CH$_3$OH and H$_2$CO, the extended emission is excitation energy dependent as mentioned in Section 5.1. The extended emission of H$_2$CO is most highly correlated with C$^{34}$S, followed closely by SO and CH$_3$CN.

For the case of HC$_3$N, the upper-state energy is 59 K, and the extended emission appears to potentially trace both the bipolar outflows emanating from W3(H$_2$O) and the low-velocity shock region. HC$_3$N is most correlated with C$^{34}$S.

*6.3. Absorption in the Dusty Cocoon*

Previous studies have detected the presence of a "dusty cocoon" surrounding the H II region in W3(OH) (Dreher & Welch 1981; Dickel et al. 1984; Qin et al. 2016). This dusty cocoon likely formed from a shell of gas and dust that traces the edge of the H II region and is kept from falling onto the central star by radiative pressure (Dreher & Welch 1981). In the current work, the spectral lines of $^{33}$SO, CS, C$^{34}$S, DCN, and H$_2$CO display P Cygni line profiles, where a broad emission line is coupled with a blueshifted absorption feature (Cid Fernandes 1999). The presence of this blueshifted absorption points toward an expanding shell around W3(OH) (Dreher & Welch 1981; Qin et al. 2016). A moment-0 map of the $^{33}$SO $3_3$–$2_2$ transition shows the absorption within the dusty cocoon and can be seen in Figure 9. This line profile of $^{33}$SO is similar to that seen by Hakobian & Crutcher (2011) for HCO$^+$. The observed absorption and line profiles indicate a layer of cold, lower-density gas surrounding the hot, dense region within the dusty cocoon (Hakobian & Crutcher 2011).

## 7. Conclusions

Here we have presented our findings from the analysis of NOEMA observations of W3(H$_2$O) and W3(OH). We fit broadband spectra using GOBASIC to identify over 20 molecules in both cores. Generally, our results agree with previous work and line analysis on W3(H$_2$O) and W3(OH) (Qin et al. 2015; Widicus Weaver et al. 2017; Ahmadi et al. 2018). For the detected molecules with a sufficient number of detected lines, we derived values for column density, line width, temperature, and velocity shift. Using the spectrum from each image pixel, we created temperature and column density maps for CH$_3$OH. We also showcase the first detections of CH$_3$CH$_2$CN and HDO in W3(OH). Previous work has reported the chemistry of W3(H$_2$O) to be more complex than that of W3(OH) (Helmich et al. 1994; Qin et al. 2015). However, in the current observations, all molecules detected in W3(H$_2$O) were also detected in W3(OH), albeit in lower abundances. These lower abundances, particularly for HDO and CH$_3$CH$_2$CN, can be explained by the destruction of the molecules in the more evolved core, which is commonly observed in H II regions (Elitzur & de Jong 1978; Rodríguez-Franco et al. 1998). These findings demonstrate the impact of physical environment on the observed chemistry in star-forming regions at different evolutionary stages.

We have also created moment-0 maps of 18 molecules to determine the spatial distributions for each of these molecules. Using the moment-0 maps, we calculated Pearson correlation coefficients to compare the spatial distribution of emission between molecules and discussed the chemical implications of highly correlated molecules. These maps also show clear evidence of a dusty cocoon within W3(OH). The images of absorption within the dusty cocoon support the idea of warm gas within the cocoon, surrounded by an envelope of cold gas. The images also show extended emission southwest of W3(OH) for several molecules; this extended emission was previously only seen for $HCO^+$ and $CH_3OH$. Based on the moment-0 maps and correlation analysis, the extended emission could potentially be indicative of a low-velocity shock in this region.

## Acknowledgments

S.L.W.W., W.E.T., and M.M.G. thank the University of Wisconsin−Madison for S.L.W.W.'s startup support and access to NOEMA time that enabled this research. We thank Ka Tat Wong from IRAM for support in setting up the observations and initial data reduction. We also thank Leon Trapman for helpful conversations pertaining to data analysis and interpretation. Part of this research was carried out at the Jet Propulsion Laboratory, California Institute of Technology, under a contract with the National Aeronautics and Space Administration (grant No. 80NM0018D0004).

*Software:* Astropy (Astropy Collaboration et al. 2013, 2018, 2022).

## ORCID iDs

Dariusz C. Lis https://orcid.org/0000-0002-0500-4700
Susanna L. Widicus Weaver https://orcid.org/0000-0001-6015-3429